\documentclass[11pt,eqs]{article}

\usepackage{latexsym,amsmath,amstext,amssymb,mathtools}
\usepackage{graphicx}
\usepackage[all]{xy}
\usepackage{tikz}
  \usetikzlibrary{shapes,arrows,fit,calc,positioning}
  \tikzset{box/.style={draw, rectangle, thick, text centered, minimum height=3em}}
  \tikzset{line/.style={draw, thick, -latex'}}

\usepackage[margin=1in]{geometry}

\title{
Courant bracket found out to be T-dual to Roytenberg bracket
\thanks{Work supported in part by
the Serbian Ministry of Education and Science, under contract No. 171031.}
}
\author{I. Ivani\v sevi\'c\thanks{e-mail: ivanisevic@ipb.ac.rs}, Lj. Davidovi\' c \thanks{e-mail: ljubica@ipb.ac.rs}, B. Sazdovi\' c \thanks{e-mail: sazdovic@ipb.ac.rs}\\
{\it Institute of Physics,}
{\it University of Belgrade,}
{\it 11080 Belgrade, Serbia}
}

\begin{document}
\maketitle

\begin{abstract}
Bosonic string moving in coordinate dependent background fields is considered. We calculate the generalized currents Poisson bracket algebra and find that it gives rise to the Courant bracket, twisted by a 2-form $2B_{\mu \nu}$. Furthermore, we consider the T-dual generalized currents and obtain their Poisson bracket algebra. It gives rise to the Roytenberg bracket, equivalent to the Courant bracket twisted by a bi-vector $\Pi^{\mu \nu}$, in case of $\Pi^{\mu \nu} = 2 {^\star B}^{\mu \nu} = \kappa \theta^{\mu \nu}$. We conclude that the twisted Courant and Roytenberg brackets are T-dual, when the quantities used for their deformations are mutually T-dual.

\end{abstract}

\section{Introduction}
Non-geometric backgrounds \cite{nongeom, nongeom1, nongeom2} include various dualities. Duality symmetry is a way to show the equivalence between two apparently different theories. Specifically, T-duality \cite{tdual, tdual1} is a symmetry between two theories corresponding to different geometries and topologies. It was firstly noticed as the spectrum equivalence of the bosonic closed string with one dimension compactified to a radius $R$, with the bosonic closed string with one dimension compactified to a radius $\alpha^\prime/R$.

The Courant bracket \cite{courant, courant1} is the generalization of the Lie bracket so that it includes both vectors and 1-forms. It is a fundamental structure of the generalized complex geometry. Vectors and 1-forms are treated on equal footing in the generalized complex structures. Many for string theory relevant geometries, such as complex, symplectic and K{\"a}hler geometry, are integrated into the framework of generalized complex structures. Moreover, the generalized complex geometry provides a framework for a unified description of diffeomorphisms and gauge transformations of the Kalb-Ramond field. Hitchin was the first one to introduce the generalized Calabi-Yau manifolds, that unified the concept of a Calabi-Yau manifold with the one of a symplectic manifold \cite{gcy}. Gualtieri in his PhD thesis contributed further to the mathematical development of generalized complex geometry \cite{gualtieri}. 

In generalized complex geometry, closure under the Courant bracket represents the integrability condition, in a same way that closure under the Lie bracket represents the integrability condition of almost complex structures. Moreover, the Courant bracket governs the gauge transformation in the double field string theory \cite{doucou}.  

The Roytenberg bracket is the generalization of the Courant bracket, so that it includes a bi-vector. It was firstly introduced by Roytenberg \cite{royt}. In \cite{nick1}, the $\sigma$-model with both 2-form and a bi-vector was considered. The Poisson bracket algebra of the generalized currents was obtained. It has been observed that, while the current algebra is anomalous, the algebra of charges is closed and gives rise to the Roytenberg bracket. In \cite{nick2}, the Roytenberg bracket was obtained by lifting the topological sector of the first order action for the NS string $\sigma$-model to three dimensions. In \cite{rroyt}, the higher order Roytenberg bracket is realized, by twisting by a p-vector. 

In this paper, we consider the closed bosonic string moving in the coordinate dependent background fields. Generalized currents are defined as linear combinations of worldsheet basis vectors with arbitrary coordinate dependent coefficients, and their Poisson bracket algebra is calculated. We follow the work of \cite{c}, that analyzed
the most general currents of the general $\sigma$ model, where it has been shown that the algebra of most general currents gives rise to the Courant bracket, twisted by the Kalb-Ramond field. Moreover, we consider the self T-duality, that is to say T-duality realized in the same phase space. The self T-duality interchanges momenta with coordinate derivatives, as well as the background fields with their T-dual background fields. Another set of generalized currents, T-dual to the aforementioned ones, are constructed and their algebra obtained. We find that their algebra gives rise to the Roytenberg bracket obtained by twisting the Courant bracket by the T-dual of the Kalb-Ramond field. Hence, we show that the twisted Courant bracket is T-dual to the corresponding Roytenberg one, obtaining the relation that connects the mathematically relevant structures with the T-duality.

\section{Hamiltonian of the bosonic string}
\setcounter{equation}{0}

Consider the closed bosonic string in the nontrivial background defined by the symmetric metric tensor field $G_{\mu \nu}$ and the Kalb-Ramond antisymmetric tensor field $B_{\mu \nu}$, as well as the constant dilaton field $\Phi = const$. In the conformal gauge, the propagation is described by the action \cite{action, regal}
\begin{equation}\label{eq:action}
S = \int_{\Sigma} {d^2 \xi \cal{L}} = \kappa \int_{\Sigma} d^2\xi
\Big[\frac{1}{2}\eta^{\alpha\beta}G_{\mu\nu}(x)
+{\epsilon^{\alpha\beta}}B_{\mu\nu}(x)\Big]
\partial_{\alpha}x^{\mu}\partial_{\beta}x^{\nu},
\end{equation}
where integration goes over a two-dimensional world-sheet $\Sigma$ parametrized by $\xi^\alpha (\xi^0 = \tau, \xi^1 = \sigma)$ with the worldsheet metric $\eta^{\alpha \beta}$. Coordinates of the D-dimensional space-time are $x^\mu (\xi),\ \mu = 0, 1, ..., D-1$, $\epsilon^{01} = -1$ and $\kappa = \frac{1}{2 \pi \alpha^\prime}$.   

It is convenient to rewrite the action (\ref{eq:action}) using the light-cone coordinates $\xi^\pm = \xi^0 \pm \xi^1$ and derivatives $\partial_\pm = \frac{1}{2}(\partial_0 \pm \partial_1) $ as 
\begin{equation} \label{eq:act}
S = \kappa \int_{\Sigma} d^2\xi
\partial_+ x^\mu \Pi_{+\mu \nu}(x) \partial_- x^\nu, 
\end{equation}
where 
\begin{equation} \label{eq:defpi}
\Pi_{\pm \mu \nu} (x) = B_{\mu \nu} (x) \pm \frac{1}{2} G_{\mu \nu} (x).
\end{equation}

The canonical momenta are given by
\begin{equation} \label{eq:pidef}
\pi_\mu  =
 \frac{\partial{\mathcal{L}}}{\partial{\dot{x}^{\mu}}} =
\kappa G_{\mu\nu}(x) \dot{x}^{\nu}
-2\kappa B_{\mu\nu}(x)x^{\prime\nu}.
\end{equation}
The Hamiltonian is obtained in a usual way,
\begin{equation} \label{eq:hh}
{\cal H}_{C} = \pi_\mu \dot{x}^\mu - \mathcal{L} = \frac{1}{2 \kappa} \pi_\mu (G^{-1})^{\mu \nu} \pi_\nu - 2 x^{\prime \mu} B_{\mu \nu} (G^{-1})^{\nu \rho} \pi_\rho + \frac{\kappa}{2} x^{\prime \mu} G^E_{\mu \nu} x^{\prime \nu},
\end{equation}
where
\begin{equation} \label{eq:geff}
G^E_{\mu \nu} = G_{\mu \nu} - 4 (B G^{-1} B)_{\mu \nu}
\end{equation}
is the effective metric. 

Energy-momentum tensor components can be written as
\begin{equation}\label{eq:tpm}
T_\pm=\mp\frac{1}{4\kappa}(G^{-1})^{\mu\nu}j_{\pm\mu}j_{\pm\nu},
\end{equation}
where the currents $j_{\pm\mu}$ are given by
\begin{equation}\label{eq:jdef}
j_{\pm\mu} (x) = \pi_\mu+2\kappa\Pi_{\pm\mu\nu}(x)x^{\prime\nu}.
\end{equation}
In terms of the energy-momentum tensor components (\ref{eq:tpm}), the Hamiltonian is given by
\begin{equation} \label{eq:h}
{\cal H}_{C}= T_- -T_+ = \frac{1}{4\kappa}(G^{-1})^{\mu\nu}\Big[
j_{+\mu}j_{+\nu}
+j_{-\mu}j_{-\nu}
\Big].
\end{equation}
In this paper, we are interested in these currents, currents T-dual to them, their generalizations, as well as their Poisson bracket algebra. Before that, let us present a short overview of T-duality.

\subsection{Lagrangian approach to T-duality}

In the Lagrangian approach to T-duality, the Buscher procedure of T-dualization has been developed \cite{buscher, buscher1, buscher2}. It provides us with the procedure of transforming coordinates from one theory to the coordinates from its T-dual theory, when there is a global Abelian isometry of coordinates along which T-dualization is applied. The T-dualization rules for coordinates  are given by \cite{tdualperm1, tdualperm2} 
\begin{equation} \label{eq:xydual}
\partial_{\pm} x^\mu \cong - \kappa \Theta^{\mu \nu}_\pm \partial_\pm y_\nu, \ \  \partial_\pm y_\mu \cong - 2 \Pi_{\mp \mu \nu} \partial_\pm x^\nu \, ,
\end{equation}
where we have introduced the T-dual coordinate $y_\mu$  and new fields $\Theta^{\mu \nu}_\pm$, defined by
\begin{equation} \label{eq:thetadef}
\Theta^{\mu \nu}_\pm = - \frac{2}{\kappa} (G_E^{-1} \Pi_\pm G^{-1})^{\mu \nu} = \theta^{\mu \nu} \mp \frac{1}{\kappa}(G_E^{-1})^{\mu \nu}, \, 
\end{equation}
where $\theta^{\mu \nu}$ is the non-commutativity parameter, that first appeared in the context of open string coordinates non-commutativity in the presence of non-zero Kalb Ramond field \cite{witten}, given by
\begin{equation} \label{eq:ncdef}
\theta^{\mu \nu} = - \frac{2}{\kappa} (G_E^{-1} B G^{-1})^{\mu \nu} \, ,
\end{equation}
where $(G_E^{-1})^{\mu \nu}$ is the inverse of the effective metric defined in (\ref{eq:geff}). It is straightforward to verify that $\Theta_\pm^{\mu \nu}$ fields are inverse to $\Pi_{\mp \mu \nu}$ fields 
\begin{equation} \label{eq:thetapi}
\Theta^{\mu \rho}_\pm \Pi_{\mp \rho \nu } = \frac{1}{2 \kappa} \delta^\mu_{\ \nu} \, .
\end{equation}
Successive application of the T-dualization (\ref{eq:xydual}) is involutive
\begin{equation}
\partial_{\pm} x^\mu \cong - \kappa \Theta^{\mu \nu}_\pm \partial_\pm y_\nu \cong 2 \kappa \Theta^{\mu \nu}_\pm \Pi_{\mp \nu \rho} \partial_{\pm} x^\rho = \partial_{\pm} x^\mu \, ,
\end{equation}
where in the last step we have used (\ref{eq:thetapi}).

Applying the T-dualization laws (\ref{eq:xydual}) to the action (\ref{eq:act}), we obtain the T-dual action
\begin{equation} \label{eq:actdual}
{^\star S} = \int d^2 \xi \ {^\star {\cal {L}}} = \frac{\kappa^2}{2} \int d^2 \xi \partial_+ y_\mu \Theta^{\mu \nu}_- \partial_- y_\nu \, .
\end{equation}
Expressing the action (\ref{eq:actdual}) in the form of the initial action (\ref{eq:act}), we obtain
\begin{equation} \label{eq:thetadualpi}
{^\star \Pi}_+^{\mu \nu} = \frac{\kappa}{2} \Theta_-^{\mu \nu} \, ,
\end{equation}
which allows us to read the T-dual background fields
\begin{equation} \label{eq:gbcan}
^\star G^{\mu \nu} = (G_E^{-1})^{\mu \nu}, \ \ \ ^\star B^{\mu \nu} = \frac{\kappa}{2} \theta^{\mu \nu} \, .
\end{equation}
These relations correspond exactly to the expressions for T-dual background fields obtained by Buscher \cite{buscher, buscher1, buscher2} in case of the existence of Abelian group of isometries along coordinates along which we perform T-duality.

\subsection{Hamiltonian formulation of T-duality}

Let us rewrite the T-dualization laws (\ref{eq:xydual}) in terms of phase space variables. Firstly, we need the expression for the T-dual canonical momentum. It is given by
\begin{equation} \label{eq:pidual}
^\star \pi^\mu = \frac{\partial \ {^\star {\cal {L}}}}{\partial \dot{y}_\mu} = \kappa (G_E^{-1})^{\mu \nu} \dot{y}_\nu - \kappa^2 \theta^{\mu \nu} y^\prime_\nu \, .
\end{equation}
Secondly, let us rewrite equations (\ref{eq:xydual}), separating the part that changes the sign from the part that does not. For the coordinates of the initial theory, we obtain
\begin{equation} \label{eq:xxydual}
\dot{x}^{\mu} \cong - \kappa \theta^{\mu \nu} \dot{y}_\nu +  (G_E^{-1})^{\mu \nu} y^\prime_\nu, \ \ \ x^{\prime \mu} \cong (G_E^{-1})^{\mu \nu}\dot{y}_\nu - \kappa \theta^{\mu \nu}  y^\prime_\nu \, , 
\end{equation}
and for the coordinates of the T-dual theory, we obtain
\begin{equation} \label{eq:yyxdual}
\dot{y}_{\mu} \cong - 2 B_{\mu \nu} \dot{x}^\nu +  G_{\mu \nu} x^{\prime \nu}, \ \ \ y^\prime_\mu \cong G_{\mu \nu}\dot{x}^\nu - 2B_{\mu \nu}  x^{\prime \nu} \, .
\end{equation}

Comparing the second relation of (\ref{eq:xxydual}) with (\ref{eq:pidef}), as well as the second relation of (\ref{eq:yyxdual}) with (\ref{eq:pidual}), we obtain the T-dualization laws (\ref{eq:xydual}) formulated in terms of the phase space variables
\begin{equation} \label{eq:tdual}
\kappa x^{\prime \mu} \cong {^\star \pi}^\mu, \ \ \ \pi_\mu \cong \kappa y^{\prime}_{\mu} \, .
\end{equation}
When coordinate $\sigma$-derivatives and canonical momenta are integrated over the worldsheet space parameter $\sigma$, the winding numbers and momenta are respectively obtained \cite{tdwcb}. Hence, we see that the T-dualization transforms the momenta of the initial theory into the winding numbers in its T-dual theory, and vice versa. 

The T-duality can be considered as the canonical transformation generated by the type I functional \cite{dualcan1, dualcan}
\begin{equation} \label{eq:genf}
F = \kappa \int d \sigma x^\mu y^\prime_\mu \, ,
\end{equation}
which gives rise to momenta
\begin{equation} \label{eq:pixveze}
\pi_\mu = \frac{\delta F}{\delta x^\mu} = \kappa y^\prime_\mu, \ \ ^\star \pi^\mu = \frac{- \delta F}{\delta y_\mu} = \kappa x^{\prime \mu} \, .
\end{equation}
This is exactly the relation (\ref{eq:tdual}). The T-duality does not change the Hamiltonian, since the generating function (\ref{eq:genf}) does not depend explicitly on time ${\cal{H}}_C \to {\cal{H}}_C + \frac{\partial F}{\partial t} = \cal{H}_C$. 

In order to obtain the T-dual Hamiltonian, we apply relations (\ref{eq:tdual}) to (\ref{eq:hh}), and obtain
\begin{equation}\label{eq:hcandual}
^\star {\cal{H}}_C =  \frac{1}{2\kappa} {^\star \pi^\mu}\  G^{E}_{\mu \nu}\  {^\star \pi^\nu}- 2\ {^\star \pi^\mu}(BG^{-1})_\mu^{\ \nu} y^\prime_\nu + \frac{\kappa}{2} y^\prime_\mu (G^{-1})^{\mu \nu} y^\prime_\nu   \, .
\end{equation}
Expressing the T-dual Hamiltonian in the form of the initial one (\ref{eq:hh}), as
\begin{equation} \label{eq:hdual}
^\star {\cal{H}}_C = \frac{1}{2\kappa} {^\star \pi^\mu}\  ^\star G^{-1}_{\mu \nu}\  {^\star \pi^\nu} - 2 y^\prime_\mu (^\star B\ ^\star G^{-1})^\mu_{\ \nu}\  {^\star \pi^\nu} + \frac{\kappa}{2} y^\prime_\mu {^\star G}_E^{\mu \nu} y^\prime_\nu  \, .
\end{equation}
We are able to read once again the expressions for the T-dual  background fields (\ref{eq:gbcan}). 

Given that we were able to write the Hamiltonian in terms of currents  $j_{\pm \mu}$, we would like to write the T-dual Hamiltonian (\ref{eq:hdual}) in terms of  T-dual currents. By analogy with the initial theory (\ref{eq:tpm}), we write the T-dual energy momentum tensor components as
\begin{equation} \label{eq:Tpmdual}
{^\star T}_\pm = \mp \frac{1}{4\kappa} {^\star G_{\mu \nu}^{-1}}\ {^\star j^\mu_\pm}\ {^\star j^\nu_\pm} \, ,
\end{equation}
where $^\star j_\pm^\mu$ are T-dual currents, given by 
\begin{equation} \label{eq:jdual}
^\star j^\mu_\pm  =  {^\star \pi}^\mu + 2 \kappa {^\star \Pi_\pm^{\mu \nu}} y^\prime_\nu \, .
\end{equation}
The T-dual Hamiltonian is then given by
\begin{equation} \label{eq:hjjdual}
^\star {\cal{H}}_C = {^\star T}_- - {^\star T}_+ =   \frac{1}{4 \kappa} {^\star G^{-1}_{\mu \nu}} \Big( {^\star j_+^\mu} {^\star j_+^\nu} + {^\star j_-^\mu} {^\star j_-^\nu} \Big),
\end{equation}
We can check that substituting (\ref{eq:jdual}) into (\ref{eq:hjjdual}), the T-dual Hamiltonian in the form (\ref{eq:hdual}) is obtained. Therefore,
\begin{equation} \label{eq:HHTT}
{\cal H}_C \cong {\cal ^\star H}_C, \ \ \ T_\pm \cong {^\star T}_\pm \, .
\end{equation}

\subsection{T-dual currents}

Let us consider the transformation of the currents under T-duality. Applying (\ref{eq:tdual}) to (\ref{eq:jdef}), we obtain
\begin{equation} \label{eq:jcan}
j_{\pm \mu}  \cong \kappa y^\prime_\mu + 2 \Pi_{\pm \mu \nu} {^\star \pi}^\nu = 2 \Pi_{\pm \mu \nu} {^\star j}_\pm^\nu \, ,
\end{equation}
where we have used (\ref{eq:thetapi}). 
Similarly, the T-dualization applied on the T-dual currents is as easily obtained
\begin{equation} \label{eq:jdualcan}
^\star j_{\pm}^{\mu} \cong \kappa x^{\prime \mu} + \kappa \Theta^{\mu \nu}_\mp \pi_\nu = \kappa \Theta^{\mu \nu}_\mp j_{\pm \nu}   \, . 
\end{equation}
The successive application of T-dualization on any current returns exactly that current. 

Although the initial and T-dual theories are equivalent (\ref{eq:HHTT}), the currents $j_{\pm \mu}$ and ${^\star j}_\pm^\mu$ do not transform exactly one into another by the T-dualization laws (\ref{eq:tdual}). There are couple of ways to see the nature of this fact. Firstly, the current $j_{\pm \mu}$ has the lower indices, while $^\star j^\mu_\pm$ has the upper indices. 

Secondly, substituting (\ref{eq:jcan}) into (\ref{eq:tpm}), we obtain the T-dual transformation of the energy momentum tensor 
\begin{equation} \label{eq:tpmtpm}
T_\pm \cong \pm \frac{1}{\kappa} {^\star j_\pm^\mu} (\Pi_{\mp}  G^{-1} \Pi_{\pm})_{\mu \nu} {^\star j_\pm^\nu} = \mp \frac{1}{4\kappa}  {^\star j}^\mu_\pm G^E_{\mu \nu} {^\star j}^\nu_\pm = {^\star T}_\pm \, ,
\end{equation}
where in the second step we used (\ref{eq:defpi}) and (\ref{eq:geff}). The direct transformation of currents under T-duality $j_{\pm \mu} \cong {^\star j}^\mu_\pm$ would violate invariance of the energy momentum tensor. The effective metric $G^E_{\mu \nu}$ in the expression for T-dual energy momentum tensor is obtained from $-\Pi_\mp G^{-1} \Pi_\pm =  \frac{1}{4}G_E$, which is only possible due to the non-trivial T-duality relation between currents (\ref{eq:jcan}).

Lastly, let us rewrite the expressions for currents in terms of coordinates, by substituting (\ref{eq:pidef}) into (\ref{eq:jdef}) and (\ref{eq:pidual}) into (\ref{eq:jdual})
\begin{equation}
j_{\pm \mu} =\kappa G_{\mu \nu} \partial_{\pm} x^\nu, \ \ \ {^\star j}^\mu_\pm = \kappa (G_E^{-1})^{\mu \nu} \partial_\pm y_\nu \, .
\end{equation}
Hence, in the same way that coordinates $\partial_\pm x^\mu$ do not transform into T-dual coordinates $\partial_\pm y_\mu$ under (\ref{eq:xydual}), in the same way the currents $j_{\pm \mu}$ do not transform into T-dual currents $^\star j_\pm^\mu$. The transformation of variables under T-duality (\ref{eq:tdual}) is presented in the Table 1.

\begin{table}[h]
{
\begin{center}
\begin{tabular}{|l|l|l|}
\hline
{\bf Initial theory}  && {\bf T-dual theory} \\ \hline
$\pi_\mu$ &$\cong$& $\kappa y^\prime_\mu$ \\
$\kappa x^{\prime \mu}$ &$\cong$& ${^\star \pi}^\mu$ \\ \hline
$j_{\pm \mu}$ &$\cong$& $2 \Pi_{\pm \mu \nu} {^\star j}^\nu_\pm$ \\
$\kappa \Theta_\mp^{\mu \nu} j_{\pm \nu}$ &$\cong$& $ {^\star j}^\mu_\pm$ \\ \hline
\end{tabular}
\end{center}
}
\caption{Transformations under the T-dualization} 
\end{table} 

Lastly, let us define for future convenience the right hand side of (\ref{eq:jdualcan}), as a new current $l_\pm^\mu$ 
\begin{equation} \label{eq:ldef}
l_\pm^\mu =  \kappa \Theta_\mp^{\mu \nu} j_{\pm \nu} = \kappa x^{\prime \mu} + \kappa \Theta_\mp^{\mu \nu} \pi_\nu \, .
\end{equation}
In the next chapter, we will see how we can avoid working in two phase spaces, and the currents $l_\pm^\mu$ will have an important role throughout the rest of the paper. 

\subsection{Self T-duality}

So far we considered the case when two mutually T-dual theories are defined in two different phase spaces, that we have marked by $\{ x^\mu, \pi_\mu \}$, and $\{y_\mu, {^\star \pi}^\mu \}$.  It is in fact possible to realize T-duality in the same phase space, that we will call self T-duality. 

To realize self T-duality, let us rewrite the second relation of (\ref{eq:xxydual}), using (\ref{eq:gbcan})
\begin{equation} \label{eq:selff}
\kappa x^{\prime \mu} \cong \kappa\ {^\star G}^{\mu \nu}\dot{y}_\nu - 2\kappa\ {^\star B}^{\mu \nu}  y^\prime_\nu \, .
\end{equation}
Comparing it with the expression for momenta (\ref{eq:pidef}), we conclude that the exchange of coordinate with its T-dual $x^\mu \leftrightarrow y_\mu$ is equivalent to
\begin{equation} \label{eq:can}
\pi_\mu \leftrightarrow \kappa x^{\prime \mu}, \ \ B_{\mu \nu} \leftrightarrow {^\star B}^{\mu \nu} = \frac{\kappa}{2} \theta^{\mu \nu},\  \ G_{\mu \nu} \leftrightarrow {^\star G}^{\mu \nu} = (G_E^{-1})^{\mu \nu}\, .
\end{equation}
These are transformation rules for what we call self T-duality. Note that unlike in (\ref{eq:tdual}), here the background fields are transformed, as well. 

The self T-duality gives the same expressions for T-dual background fields (\ref{eq:gbcan}) as in case of Buscher procedure. It swaps the winding numbers with momenta as well, therefore preserving all features of T-duality, with the only difference being that it is realized in the same phase space.

The two currents $j_{\pm \mu}$ and $l_\pm^\mu$ transform into each other under the self T-duality (\ref{eq:can}) 
\begin{equation} \label{eq:jl}
j_{\pm \mu} = \pi_\mu + 2\kappa \Pi_{\pm \mu \nu} x^{\prime \nu} \leftrightarrow  \kappa x^{\prime \mu} + \kappa \Theta^{\mu \nu}_\mp \pi_\nu = l_\pm^\mu \,  .
\end{equation}
On the other hand, under (\ref{eq:can}) the energy-momentum tensor is invariant
\begin{equation} \label{eq:Tsd}
T_\pm= \mp\frac{1}{4\kappa}(G^{-1})^{\mu\nu}j_{\pm\mu}j_{\pm\nu}  \leftrightarrow \mp\frac{1}{4\kappa}G^E_{\mu\nu}l_\pm^\mu l_\pm^\nu = T_\pm \, .
\end{equation}
With the help of (\ref{eq:h}), we see that the Hamiltonian does not change under (\ref{eq:can}). Nevertheless, the Hamiltonian can be expressed in terms of new currents $l_\pm^\mu$ 
\begin{equation} \label{eq:Hll}
 {\cal H}_C = \frac{1}{4 \kappa} G^E_{\mu \nu} \Big( l_+^\mu l_+^\nu + l_-^\mu l_-^\nu \Big) \, ,
\end{equation}
but with the effective metric instead of the inverse metric. Substituting (\ref{eq:geff}) and (\ref{eq:ldef}) into the previous equation, we obtain the initial form of the Hamiltonian (\ref{eq:h}).

It is important to point out that although the energy-momentum tensor components $T_\pm$ and the Hamiltonian ${\cal H}_C$ remain invariant under the self T-duality, the currents $j_{\pm \mu}$ and $l_\pm^\mu$ do not. Therefore although both currents $j_{\pm \mu}$ and $l_\pm^\mu$ are defined in terms of the initial theory variables, they have to change under self T-duality, due to the invariance of  energy momentum tensor components (\ref{eq:Tsd}). We summarize its transformation rules in the Table 2. Our next goal is to generalize these two currents and obtain the algebra of their generalizations.  

\begin{table}[h]
{
\begin{center}
\begin{tabular}{|l|l|l|}
\hline
{\bf Initial theory}  && {\bf Self T-dual theory} \\ \hline
$\pi_\mu$ &$\leftrightarrow$& $\kappa x^\prime_\mu$ \\
$\kappa x^{\prime \mu}$ &$\leftrightarrow$& $\pi_\mu$ \\ \hline
$ B_{\mu \nu}$ & $\leftrightarrow$ & $ \frac{\kappa}{2} \theta^{\mu \nu}$ \\
$G_{\mu \nu}$ &$\leftrightarrow$& $(G_E^{-1})^{\mu \nu}$ \\  \hline
$j_{\pm \mu}$ &$\leftrightarrow$& $l^\mu_\pm$ \\ \hline
\end{tabular}
\end{center}
}
\caption{Transformations under the self T-duality} 
\end{table} 

\section{Generalized currents in a new basis}
\setcounter{equation}{0}

In this chapter, we will construct two types of generalized currents. Generalized currents are arbitrary functionals of the fields, parametrized by a pair of vector field
and covector field on the target space, treating both vectors and 1-forms on equal footing \cite{gualtieri}. The convenient bases in which these generalized currents are defined are components of currents $j_{\pm \mu}$ and $l_\pm^\mu$. 

Firstly, we will generalize the currents $j_{\pm \mu}$. From (\ref{eq:jdef}) we extract its $\tau$ and $\sigma$ components
\begin{equation}
j_{0 \mu} = \frac{j_{+ \mu} + j_{- \mu}}{2} = \pi_{\mu} + 2 \kappa B_{\mu \nu} (x) x^{\prime \nu},\ \ \ j_{1 \mu} = \frac{j_{+ \mu} - j_{- \mu}}{2} = \kappa G_{\mu \nu} (x) x^{\prime \nu}.
\end{equation}
We will mark 
\begin{equation} \label{eq:idef}
i_\mu = \pi_{\mu} + 2 \kappa B_{\mu \nu} (x) x^{\prime \nu},
\end{equation}
 as a new, auxiliary current. Therefore, $\{\kappa x^{\prime \mu},i_\mu \}$ is a new convenient basis on the world-sheet. We can now write currents (\ref{eq:jdef}) in this basis as
\begin{equation} \label{eq:jiveza}
j_{\pm \mu} = i_\mu \pm \kappa G_{\mu \nu} x^{\prime \nu}.
\end{equation}

In the same way as in \cite{c}, we define the generalized currents in the new basis, as the linear combination of both coordinate $\sigma-$derivatives and auxiliary currents
\begin{equation}\label{eq:JCdef}
J_{C(u, a)} = u^\mu (x) i_\mu + a_\mu (x) \kappa x^{\prime \mu},
\end{equation}
where $u^\mu (x)$ and $a_\mu (x)$ are the arbitrary coefficients. The charges of these currents are
\begin{equation}\label{eq:QCdef}
Q_{C(u, a)} = \int d \sigma J_{C(u, a)}. 
\end{equation}

The charges exhibit additional symmetry. In order to see that, let us firstly rewrite the integral of the total derivative of an arbitrary function $\lambda$  
\begin{equation} \label{eq:tdlambda}
 \int_0^{2\pi} d\sigma (\lambda)^{\prime} = \int_0^{2\pi} d\sigma \partial_\mu \lambda x^{\prime \mu}= 0 \, ,
\end{equation}
which goes to zero for closed strings.
From this fact, we obtain the reducibility relations for the charges 
\begin{equation} \label{eq:QCred}
Q_{C(u,a+\partial \lambda)} = Q_{C(u,a)} \, .
\end{equation}

The expression of the form (\ref{eq:JCdef}) is particularly interesting, since it gives rise to many for string theory relevant structures.  Firstly, for the special case of coefficients relation $a_\mu = \pm  G_{\mu \nu} u^\nu$, we obtain
\begin{equation} \label{eq:jcj}
J_{C(u, \pm  G u)} = u^\mu j_{\pm \mu}. 
\end{equation}
Hence, the currents (\ref{eq:jdef}) indeed can be obtained from the generalized currents (\ref{eq:JCdef}).
On the other hand, for special case $a_\mu = - 2  B_{\mu \nu} u^\nu$, we obtain 
\begin{equation}
J_{C(u,-2 Bu)} = u^\mu \pi_\mu,
\end{equation}
as well as for $u^\mu =0$, we obtain 
\begin{equation}
J_{C(0,a)} = a_\mu \kappa x^{\prime \nu}.
\end{equation}
We see that the general current algebra for the appropriate coefficients reduces to non-commutativity relations of both coordinates and momenta.

We are also interested in another type of generalized current, that in analogous way generalizes $l_\pm^\mu$, in the basis related to its $\tau$ and $\sigma$ components
\begin{equation}
l_0^\mu = \frac{l_+^\mu + l_-^\mu}{2} = \kappa x^{\prime \mu} + \kappa \theta^{\mu \nu} \pi_\nu,\ \ \ l_1^\mu = \frac{l_+^\mu - l_-^\mu}{2} =  (G_E^{-1})^{\mu \nu}  \pi_\nu.
\end{equation}
The second set of generalized currents is defined by
\begin{equation} \label{eq:JRdef}
J_{R(v,b)} = v^\mu (x) \pi_\mu + b_\mu(x) k^\mu,  
\end{equation}
where $v^\mu(x)$ and $b_\mu (x)$ are the arbitrary coefficients, and we have introduced another auxiliary current by
\begin{equation} \label{eq:kdef}
k^\mu =  \kappa x^{\prime \mu} + \kappa \theta^{\mu \nu} \pi_\nu .
\end{equation}
Their charges are
\begin{equation} \label{eq:QRdef}
Q_{R(v,b)} = \int d \sigma J_{R(v,b)}.
\end{equation}
Similarly as in (\ref{eq:QCred}), these charges also exhibit additional symmetry. In order to see that, let us write the total derivative integral (\ref{eq:tdlambda}), using (\ref{eq:kdef}), in terms of new basis vectors 
\begin{equation}
\int_0^{2\pi} d\sigma \kappa \partial_\mu \lambda x^{\prime \mu} = \int_0^{2\pi} d\sigma \partial_\mu \lambda (k^\mu -\kappa \theta^{\mu\nu} \pi_\nu) \, .
\end{equation}
As a result, we obtain the non-uniqueness of the charges
\begin{equation} \label{eq:QRred}
Q_{R(v+\kappa \theta \partial \lambda,b + \partial \lambda)}  = Q_{R(v,b)} \, .
\end{equation}

In a special case of $v^\mu = \pm (G_E^{-1})^{\mu \nu} b_\nu$, the generalized current (\ref{eq:JRdef}) reduces to the current (\ref{eq:jcan})
\begin{equation}
J_{R(\pm G_E^{-1} b,b)} = b_\mu l^\mu_{\pm}, 
\end{equation}
thus justifying calling it generalized current. Momenta $\pi_\mu$ and auxiliary currents $k^\mu$ can also be as easily obtained from it. 

The two new bases transform into each other under (\ref{eq:can}):
\begin{equation}
i_\mu = \pi_\mu + 2\kappa B_{\mu \nu} x^{\prime \nu} \leftrightarrow \kappa x^{\prime \mu} + \kappa \theta^{\mu \nu} \pi_\nu = k^\mu, \ \ \pi_\mu \leftrightarrow \kappa x^{\prime \mu} \, .
\end{equation}
Therefore, the generalized currents are defined in the mutually T-dual bases, and their respective algebras are also going to be mutually T-dual. 

At the end of this chapter, let us obtain the relations for coefficients when two generalized currents are equal. This will enable us to obtain the algebra of currents $J_{R(v,b)}$, provided that we have the algebra of $J_{C(u,a)}$, and vice versa. Let us start with rewriting the expressions for both generalized currents in the basis $\{\pi_\mu, x^{\prime \mu} \}$. Substituting the expression (\ref{eq:idef}) into (\ref{eq:JCdef}) we obtain
\begin{equation} \label{eq:JCB}
J_{C(u,a)} = u^\mu \pi_\mu + \kappa( a_\mu - 2 B_{\mu \nu} u^{\nu}) x^{\prime \mu},
\end{equation}
while substituting the expression (\ref{eq:kdef}) into (\ref{eq:JRdef}) we obtain
\begin{equation} \label{eq:JRB}
J_{R(v,b)} = (v^\mu - \kappa \theta^{\mu \nu} b_\nu) \pi_\mu + \kappa b_\mu x^{\prime \mu}.
\end{equation}
Comparing (\ref{eq:JCB}) to (\ref{eq:JRB}), we see that generalized currents are equal when coefficients satisfy following relations
\begin{align} \label{eq:coeffc}
u^\mu &= v^\mu - \kappa \theta^{\mu \nu} b_\nu, \\ \notag
a_\mu &= 2B_{\mu \nu} v^\nu +  (G G_E^{-1})_\mu^{\ \nu} b_\nu \, .
\end{align}
The above relations can be easily inversed. We obtain
\begin{align} \label{eq:coeffr}
v^\mu &= (G_E^{-1} G)^\mu_{\ \nu} u^\nu + \kappa \theta^{\mu \nu} a_\nu, \\ \notag
 b_\mu &= a_\mu - 2B_{\mu \nu} u^\nu. 
\end{align}

\section{Courant bracket}
\setcounter{equation}{0}
We are interested in calculating the Poisson bracket algebra of the most general currents $J_{C(u,a)}$, defined in (\ref{eq:JCdef}), as well as of their charges $Q_{C(u,a)}$, defined in (\ref{eq:QCdef}). We will start with the generators $i_\mu$ and $x^{\prime \mu}$ algebra, that we calculate using the standard Poisson bracket relations 
\begin{align} \label{eq:pbr} \notag
&\{ x^\mu (\sigma, \tau), \pi_\nu (\bar{\sigma}, \tau) \} = \delta^\mu_{\ \nu} \delta(\sigma-\bar{\sigma}), \\ 
& \{ x^\mu (\sigma, \tau), x^\nu (\bar{\sigma}, \tau) \} = 0, \\ \notag 
& \{ \pi_\mu (\sigma, \tau), \pi_\nu (\bar{\sigma}, \tau) \} = 0.
\end{align} 
In the accordance with \cite{c}, we will obtain that the algebra of generalized charges (\ref{eq:QCdef}) gives rise to the twisted Courant bracket \cite{courant}. 

We obtain the algebra of generators (\ref{eq:idef})
\begin{equation} \label{eq:pbri}
\{ i_\mu (\sigma), i_\nu (\bar{\sigma}) \} = - 2\kappa B_{\mu \nu \rho} x^{\prime \rho} \delta (\sigma - \bar{\sigma}),
\end{equation}
where the structural constants are the Kalb-Ramond field strength components, given by 
\begin{equation} \label{eq:bmnr}
B_{\mu \nu \rho} = \partial_\mu B_{\nu \rho} + \partial_\nu B_{\rho \mu} + \partial_\rho B_{\mu \nu}.
\end{equation} 
The rest of the generators algebra is given by
 \begin{equation} \label{eq:pbrix}
\{i_\mu (\sigma), \kappa x^{\prime \nu} (\bar{\sigma}) \} = \kappa  \delta_\mu^{\ \nu} \partial_\sigma \delta(\sigma-\bar{\sigma}),\ \ \ \{ \kappa  x^{\prime \mu} (\sigma), \kappa  x^{\prime \nu} (\bar{\sigma}) \} = 0.
\end{equation}
The Poisson bracket of the most general currents (\ref{eq:JCdef}) is obtained using (\ref{eq:pbri}) and (\ref{eq:pbrix}). It reads
\begin{align} \notag \label{eq:jcalg}
\{ J_{C(u, a)} (\sigma), J_{C(v, b)}(\bar{\sigma}) \} = & (v^\nu \partial_\nu u^\mu - u^\nu \partial_\nu v^\mu) i_\mu \delta(\sigma-\bar{\sigma}) - 2 \kappa B_{\mu \nu \rho} x^{\prime \mu} u^\nu v^\rho \delta(\sigma-\bar{\sigma})  \\
& - \kappa \Big( (\partial_\mu a_\nu - \partial_\nu a_\mu) v^\nu - (\partial_\mu b_\nu - \partial_\nu b_\mu) u^\nu \Big) x^{\prime \mu}\delta(\sigma-\bar{\sigma})  \\ \notag
&+\kappa \Big(u^\mu (\sigma) b_\mu (\sigma) + v^\mu (\bar{\sigma}) a_\mu (\bar{\sigma}) \Big) \partial_\sigma \delta(\sigma-\bar{\sigma}).
\end{align}
We can modify the anomalous part in the following manner
\begin{align} \notag \label{eq:anomtrans}
 \Big( & u^\mu (\sigma)  b_\mu (\sigma) + v^\mu (\bar{\sigma}) a_\mu (\bar{\sigma}) \Big) \partial_\sigma \delta(\sigma- \bar{\sigma}) = \\
& =\frac{1}{2}\Big( (ub) (\sigma) + (va)(\bar{\sigma}) \Big) \partial_\sigma \delta(\sigma- \bar{\sigma})- \frac{1}{2} (ub) (\sigma) \partial_{\bar{\sigma}} \delta(\sigma-\bar{\sigma}) + \frac{1}{2}(va) (\bar{\sigma}) \partial_\sigma \delta(\sigma-\bar{\sigma}) \\ \notag
& = \frac{1}{2}\Big( (ub) (\sigma) + (ub)(\bar{\sigma})+ v a(\sigma) +(va)(\bar{\sigma}) \Big) \partial_\sigma \delta(\sigma- \bar{\sigma}) + \frac{1}{2}\partial_\mu (va-ub) x^{\prime \mu} \delta(\sigma-\bar{\sigma}),
\end{align}
where we have used the notation $(ub)(\sigma) = u^\mu (\sigma) b_\mu (\sigma)$, and the relation $f(\bar{\sigma}) \partial_\sigma \delta(\sigma-\bar{\sigma}) = f^\prime(\sigma) \delta(\sigma-\bar{\sigma}) +f(\sigma) \partial_\sigma \delta(\sigma-\bar{\sigma})$ in the last step. Substituting the previous equation in (\ref{eq:jcalg}) we obtain 
\begin{equation} \label{eq:pbrJC}
\{ J_{C(u, a)} (\sigma), J_{C(v, b)}(\bar{\sigma}) \} = -J_{C(\bar{w}, \bar{c})}(\sigma) \delta(\sigma-\bar{\sigma}) + \frac{\kappa}{2}\Big((ub) (\sigma) + (ub)(\bar{\sigma})+ (v a)(\sigma) +(va)(\bar{\sigma}) \Big) \partial_\sigma \delta(\sigma-\bar{\sigma}),
\end{equation}
where the coefficients in the resulting current are
\begin{equation} \label{eq:wdef}
\bar{w}^\mu =  u^\nu \partial_\nu v^\mu - v^\nu \partial_\nu u^\mu,
\end{equation}
and
\begin{equation} \label{eq:cdef}
\bar{c}_\mu = 2  B_{\mu \nu \rho} u^\nu v^\rho +  (\partial_\mu a_\nu - \partial_\nu a_\mu) v^\nu -  (\partial_\mu b_\nu - \partial_\nu b_\mu) u^\nu + \frac{1}{2} \partial_\mu (ub - va).
\end{equation}
The minus sign in front of the $J_{C(\bar{w}, \bar{c})}$ is included for the future convenience. We see that $\bar{w}^\mu$ does not depend on background fields, while the coefficient $\bar{c}_\mu$ does, because of the $H$-flux term $B_{\mu \nu \rho}$.

The relation (\ref{eq:pbrJC}) defines the bracket, that acts on a pair of two ordered pairs consisting of a vector and a 1-form, that as a result has another ordered pair, that we can write like 
\begin{equation} \label{eq:motc}
[(u,a), (v,b)]_{C} =(\bar{w},\bar{c}). 
\end{equation}
The bracket that we have obtained is the twisted Courant bracket \cite{courant}. The Courant bracket represents the generalization of the Lie bracket on spaces that contain both vectors and 1-forms. As a result, it gives an ordered pair of a vector $w = w^\mu \partial_\mu$ and a 1-form $c = c_\mu d x^\mu$.

Let us confirm the equivalence between the twisted Courant bracket and the bracket that we have obtained in (\ref{eq:motc}).
The coordinate free expression for the twisted Courant bracket is given by
\begin{equation} \label{eq:cfcourant}
[(u,a), (v,b)]_C = \Big( [u,v]_L, {\cal{L}}_u b - {\cal{L}}_v a - \frac{1}{2}d (i_u b - i_v a) + H(u,v,.)  \Big) \equiv (w,c) ,
\end{equation}
where $[u,v]_L$ is the Lie bracket  and $H(u,v,.)$ is a 1-form obtained by contracting a three form. The Lie derivative ${\cal{L}}_u$ is defined in a usual way ${\cal{L}}_u = i_u d + d i_u$, where $d$ is the exterior derivative and  $i_u$ the interior derivative. Their action on 1-forms is given by $d a = \partial_\mu a_\nu dx^\mu dx^\nu$ and $i_u a = u^\mu a_\mu$. 

The Lie bracket is given by
\begin{equation} \label{eq:cr1}
\left. [u,v]_L \  \right|^\mu = u^\nu \partial_\nu v^\mu - v^\nu \partial_\nu u^\mu. 
\end{equation}
Using the definition of Lie derivative, we furthermore obtain
\begin{equation} \label{eq:rt2}
\left. \Big( {\cal{L}}_u b - {\cal{L}}_v a - \frac{1}{2}d (i_u b - i_v a) \Big) \right|_\mu = u^\nu (\partial_\nu b_\mu - \partial_\mu b_\nu) - v^\nu(\partial_\nu a_\mu - \partial_\mu a_\nu) + \frac{1}{2} \partial_\mu (ub - va).
\end{equation}
As for the last term in (\ref{eq:cfcourant}), it is given by
\begin{equation} \label{eq:cr2}
\left. H(u,v,.)  \right|_\mu = 2  B_{\mu \nu \rho} u^\nu v^\rho.
\end{equation}
The expression for the generalized current corresponding to the Courant bracket is obtained by substituting  (\ref{eq:cr1}), (\ref{eq:rt2}) and (\ref{eq:cr2}) in (\ref{eq:cfcourant})
\begin{equation} \label{eq:courantcourant}
[(u,a),(v,b)]_C = (w,c),
\end{equation}
where $w^\mu$ and $c_\mu$ are exactly the same as $\bar{w}^\mu$ and $\bar{c}_\mu$ defined in (\ref{eq:wdef}) and (\ref{eq:cdef}), respectively. Therefore, we see that the bracket defined in (\ref{eq:motc}) is indeed the twisted Courant bracket.

Besides the current algebra, we are interested in the algebra of charges (\ref{eq:QCdef}). The anomalous term is canceled when integrated. For example, consider the first term in anomaly
\begin{equation}
\int d \sigma d \bar{\sigma} (ub) (\sigma) \partial_\sigma \delta(\sigma-\bar{\sigma}) = - \int d\bar{\sigma} \partial_{\bar{\sigma}} \int d \sigma (ub) (\sigma) \delta (\sigma - \bar{\sigma}) = - \int d\bar{\sigma} \partial_{\bar{\sigma}} (ub(\bar{\sigma})) = 0,
\end{equation}
since we are working with the closed strings. The other terms cancel in a similar manner. Integrating the generalized currents (\ref{eq:pbrJC}) over $\sigma$ and $\bar{\sigma}$ we obtain
\begin{equation} \label{eq:qqc}
\{ Q_{C(u, a)}, { Q_{C(v, b)} \}} =  - Q_{C[(u,a),(v,b)]_C}.
\end{equation}
We see that the algebra of charges is anomaly free. The relation (\ref{eq:qqc}) was firstly obtained in \cite{c} for the general case of the Hamiltonian formulation of string $\sigma$-model, in which momenta and coordinates satisfy the same Poisson bracket relations as auxiliary currents and coordinates in our theory.  

Let us check whether the algebra (\ref{eq:pbrJC}) is consistent with the known results for the Poisson bracket algebra of the currents $j_{\pm \mu}$ \cite{dualsim}
\begin{align} \label{eq:jjpbr}
&\{ j_{\pm \mu} (\sigma), j_{\pm \nu} (\bar{\sigma}) \} = \pm 2\kappa \Gamma_{\mu, \nu \rho} x^{\prime \rho} \delta(\sigma- \bar{\sigma}) - 2 \kappa B_{\mu \nu \rho} x^{\prime \rho} \delta(\sigma- \bar{\sigma}) \pm 2 \kappa G_{\mu \nu} \delta^\prime(\sigma- \bar{\sigma}), \\ \notag
&\{ j_{\pm \mu} (\sigma), j_{\mp \nu} (\bar{\sigma}) \} = \pm 2\kappa \Gamma_{\mu, \nu \rho} x^{\prime \rho} \delta(\sigma- \bar{\sigma}) - 2 \kappa B_{\mu \nu \rho} x^{\prime \rho} \delta(\sigma- \bar{\sigma}), 
\end{align}
where $\Gamma_{\mu, \nu \rho} = \frac{1}{2} (\partial_\nu G_{\rho \mu} + \partial_\rho G_{\nu \mu}-\partial_\mu G_{\nu \rho})$ are Christoffel symbols. If we substitute $a_\mu = \pm  G_{\mu \nu} u^\nu$ and $b_\mu = \pm   G_{\mu \nu} v^\nu$ for constants $u^\mu$ and $v^\mu$ in (\ref{eq:pbrJC}), with the help of (\ref{eq:jcj}) we obtain
\begin{align} \notag
\{u^\mu j_{\pm \mu}(\sigma), v^\nu j_{\pm \nu} (\bar{\sigma}) \} =& u^\mu v^\nu \Big( -2 \kappa B_{\mu \nu \rho} \pm \kappa(\partial_\nu G_{\rho \mu} - \partial_\mu G_{\nu \rho}) \Big)x^{\prime \rho} \delta(\sigma-\bar{\sigma}) \\ \notag
&\pm \kappa u^\mu v^\nu (G_{\mu \nu}(\sigma) + G_{\mu \nu}(\bar{\sigma})) \partial_\sigma \delta(\sigma-\bar{\sigma})  \\
=& u^\mu v^\nu \Big( -2 \kappa B_{\mu \nu \rho} \pm \kappa(\partial_\nu G_{\rho \mu} +\partial_\rho G_{\mu \nu} - \partial_\mu G_{\nu \rho}) \Big)x^{\prime \rho} \delta(\sigma-\bar{\sigma}) \\ \notag
&\pm 2\kappa u^\mu v^\nu G_{\mu \nu}(\sigma) \partial_\sigma \delta(\sigma-\bar{\sigma}) \\ \notag
=&  u^\mu v^\nu \{ j_{\pm \mu}, j_{\pm \nu} \}.
\end{align}
The consistency with the second relation in (\ref{eq:jjpbr}) can be as easily obtained.

\section{Roytenberg bracket}
\setcounter{equation}{0}
The Roytenberg bracket appeared as a result of the current algebra firstly in \cite{nick1}, where the author twisted the Poisson structure by trading the 2-form $B_{\mu \nu}$ with the bi-vector $\Pi^{\mu \nu}$. In this paper, we firstly calculate the Poisson bracket algebra for the generalized currents $J_{R(v,b)}$ (\ref{eq:JRdef}), in order to calculate the T-dual Poisson structure of the twisted Courant bracket.

While the currents (\ref{eq:JRdef}) have the same form as the currents giving the Roytenberg bracket in \cite{nick1}, in \cite{nick1} the momenta are redefined so that they are equal to the auxiliary currents $i_\mu$ (\ref{eq:idef}) in our paper. As a result of this difference, the currents $J_{R(v,b)}$ and $J_{C(u,a)}$ are related by self T-duality, which is not the case for corresponding currents in \cite{nick1}. Therefore, we will show that the Courant bracket twisted by a 2-form $2B_{\mu \nu}$ is T-dual to the Roytenberg bracket, obtained by twisting the Courant bracket by a bi-vector $\kappa \theta^{\mu \nu}$. When the fluxes are turned off, both of them reduce to the untwisted Courant bracket, that is T-dual to itself.

We will start with the algebra of auxiliary currents $k^\mu$  (\ref{eq:kdef}). Using (\ref{eq:pbr}), we obtain
\begin{equation}
\{ k^\mu (\sigma), k^{\nu} (\bar{\sigma}) \} = -\kappa \partial_\rho \theta^{\mu\nu} x^{\prime \rho} \delta(\sigma-\bar{\sigma}) - \kappa^2 (\theta^{\mu \sigma} \partial_\sigma \theta^{\nu \rho} - \theta^{\nu \sigma} \partial_\sigma \theta^{\mu \rho}) \pi_\rho \delta(\sigma-\bar{\sigma}),
\end{equation}
where $\theta^{\mu \nu}$ is the non-commutativity parameter (\ref{eq:ncdef}).
From (\ref{eq:kdef}) we express the coordinate in terms of algebra generators and obtain
\begin{equation} \label{eq:pbrk}
\{ k^\mu (\sigma), k^{\nu} (\bar{\sigma}) \} = -\kappa Q_\rho^{\ \mu \nu} k^\rho \delta(\sigma - \bar{\sigma}) - \kappa^2 R^{\mu \nu \rho} \pi_\rho \delta (\sigma - \bar{\sigma}),
\end{equation}
where we expressed the structure constants as fluxes
\begin{equation} \label{eq:QRdef}
Q_{\rho}^{\ \mu \nu} = \partial_\rho \theta^{\mu \nu}, \ \ \ \ \ R^{\mu \nu \rho} = \theta^{\mu \sigma} \partial_\sigma \theta^{\nu \rho} + \theta^{\nu \sigma} \partial_\sigma \theta^{\rho \mu} + \theta^{\rho \sigma} \partial_\sigma \theta^{\mu \nu}.
\end{equation}
These are the non-geometric fluxes \cite{stw}. They were firstly obtained by applying the Buscher rules \cite{buscher, buscher1, buscher2} on the three-torus with non-trivial Kalb-Ramond field strength (\ref{eq:bmnr}). After the T-duality transformations are applied along two isometry directions, one obtains the space that is locally geometric, but globally non-geometric. The flux for this background is $Q_{\rho}^{\ \mu \nu}$. After the T-duality transformation is applied along all directions, one obtains the space that is neither locally, nor globally geometric, characterized with the $R^{\mu  \nu \rho}$ flux. When considering a generalized T-dualization, the $R$ flux is obtained when performing T-dualization over the arbitrary coordinate on which the background fields depend \cite{DD}.

The rest of the generators algebra is calculated in a similar way
\begin{equation} \label{eq:pbrkpi}
\{ k^\mu (\sigma), \pi_\nu (\bar{\sigma}) \} = \kappa \delta^\mu_{\ \nu} \partial_\sigma \delta(\sigma - \bar{\sigma}) + \kappa Q_\nu^{\ \mu \rho} \pi_\rho \delta(\sigma - \bar{\sigma}),\ \ \ \{ \pi_\mu (\sigma), \pi_\nu (\bar{\sigma}) \} = 0.
\end{equation}

We obtain the Poisson bracket of the most general currents $J_{R(u,a)}$, using (\ref{eq:pbrk}) and (\ref{eq:pbrkpi}). It reads
\begin{align} \notag \label{eq:jralg}
\{ J_{R(u, a)} (\sigma), J_{R(v, b)}(\bar{\sigma}) \} = & (v^\nu \partial_\nu u^\mu - u^\nu \partial_\nu v^\mu) \pi_\mu \delta(\sigma-\bar{\sigma}) - \kappa^2 R^{\mu \nu \rho} \pi_\mu a_\nu b_\rho \delta(\sigma-\bar{\sigma})  \\ \notag
& - \kappa \Big( \theta^{\nu \rho} \partial_\rho v^\mu a_\nu -  v^\rho \partial_\nu a_\rho \theta^{\nu \mu} - \partial_\nu \theta^{\rho \mu} v^\nu a_\rho \Big) \pi_\mu \delta(\sigma-\bar{\sigma})  \\ 
& - \kappa \Big( u^\rho \partial_\nu b_\rho \theta^{\nu \mu} + \kappa u^\rho \partial_\rho \theta^{\nu \mu} b_\nu - \kappa \theta^{\nu \rho} \partial_\rho u^\mu b_\nu \Big) \pi_\mu \delta(\sigma-\bar{\sigma})  \\ \notag
& +\Big( u^\nu (\partial_\mu b_\nu - \partial_\nu b_\mu) -v^\nu (\partial_\mu a_\nu - \partial_\nu a_\mu) \Big) k^\mu \delta(\sigma-\bar{\sigma}) \\ \notag
& - \kappa \Big( a_\rho b_\nu \partial_\mu \theta^{\rho \nu} -  \theta^{\nu \rho} (\partial_\rho a_\mu b_\nu - \partial_\rho b_\mu a_\nu) \Big) k^\mu \delta(\sigma-\bar{\sigma})\\ \notag
&+\kappa \Big(u^\mu (\sigma) b_\mu (\sigma) + v^\mu (\bar{\sigma}) a_\mu (\bar{\sigma}) \Big) \partial_\sigma \delta(\sigma-\bar{\sigma}).
\end{align}
Using (\ref{eq:anomtrans}) and (\ref{eq:kdef}) we can transform the anomaly in the following way
\begin{align} \label{eq:ranom}
\kappa \Big((u b) (\sigma) + (v a) (\bar{\sigma}) \Big) \partial_\sigma \delta(\sigma-\bar{\sigma}) =& \frac{\kappa}{2}((ub) (\sigma) + (ub)(\bar{\sigma})+ (va)(\sigma) +(va)(\bar{\sigma})) \partial_\sigma \delta(\sigma- \bar{\sigma}) \\ \notag 
&+ \frac{1}{2}\partial_\mu (va - ub)(\sigma) (k^{\mu} - \theta^{\mu \rho} \pi_\rho) \delta(\sigma-\bar{\sigma}).
\end{align}
Substituting the last equation in (\ref{eq:jralg}), we obtain
\begin{equation} \label{eq:pbrJR}
\{ J_{R(u, a)} (\sigma), J_{R(v, b)}(\bar{\sigma}) \} = -J_{R(\bar{w},\bar{c})}(\sigma) \delta(\sigma-\bar{\sigma}) + \frac{\kappa}{2}\Big((ub) (\sigma) + (ub)(\bar{\sigma})+ (v a)(\sigma) +(va)(\bar{\sigma}) \Big) \partial_\sigma \delta(\sigma-\bar{\sigma}),
\end{equation}
where
\begin{align} \label{eq:defw}
\bar{w}^\mu =&\   u^\nu \partial_\nu v^\mu - v^\nu \partial_\nu u^\mu +\kappa \theta^{\nu \rho} \partial_\rho v^\mu a_\nu - \kappa v^\rho \partial_\nu a_\rho \theta^{\nu \mu} - \kappa Q_\nu^{\ \rho \mu} v^\nu a_\rho   \\ \notag
& +\kappa u^\rho \partial_\nu b_\rho \theta^{\nu \mu} + \kappa u^\rho Q_\rho^{\ \nu \mu} b_\nu - \kappa \theta^{\nu \rho} \partial_\rho u^\mu b_\nu - \frac{\kappa}{2} \theta^{\mu \nu} \partial_\nu (va - ub) + \kappa^2 R^{\mu \nu \rho} a_\nu b_\rho,
\end{align}
and
\begin{equation} \label{eq:defc}
\bar{c}_\mu = v^\nu (\partial_\mu a_\nu - \partial_\nu a_\mu) - u^\nu (\partial_\mu b_\nu - \partial_\nu b_\mu) -\frac{1}{2}\partial_\mu(va-ub)+ \kappa a_\rho b_\nu Q_\mu^{\ \rho \nu} - \kappa \theta^{\nu \rho} (\partial_\rho a_\mu b_\nu - \partial_\rho b_\mu a_\nu), 
\end{equation}
where we have substituted $Q$ and $R$ fluxes (\ref{eq:QRdef}). Unlike the coefficients in the previous case, here both coefficients depend on backgrounds, due to the presence of fluxes.

As expected, algebra is not closed due to the anomalous part. This Poisson bracket defines a new bracket 
\begin{equation} \label{eq:motr}
[(u,a),(v,b)]_{R} = (\bar{w},\bar{c}),
\end{equation}
which is equal to the Roytenberg bracket \cite{royt}. In case of only $R$ and $Q$ flux present in the generators algebra (\ref{eq:QRdef}), the Roytenberg bracket is given by
\begin{align} \label{eq:cfroytenberg}
[(u,a),(v,b)]_R = & \Big( [u,v]_L - [v,a \Pi]_L + [u,b \Pi]_L + \frac{1}{2} [\Pi,\Pi]_S(a, b, .) \\ \notag
&- \Big( {\cal{L}}_v a -{\cal{L}}_{u}b + \frac{1}{2} d (i_u b - i_v a) \Big) \Pi , \\ \notag
& + {\cal{L}}_{u}b - {\cal{L}}_v a -\frac{1}{2} d (i_u b - i_v a)  -  [a,b]_{\Pi} \Big) , \notag
\end{align}
where $\Pi = \Pi^{\mu \nu} \partial_\mu \partial_\nu$ is the bi-vector. The expression $ [\Pi,\Pi]_S(a, b, .)$ represents the Schouten-Nijenhuis bracket \cite{SNB} contracted with two 1-forms and $[a,b]_{\Pi}$ is the Koszul bracket \cite{koszul} given by
\begin{equation} \label{eq:koszul}
[a, b]_\Pi = {\cal{L}}_{a \Pi} b - {\cal{L}}_{b\Pi} a + d(\Pi(a,b)).
\end{equation}
The Koszul bracket is a generalization of the Lie bracket on the space of differential forms, while the Schouten-Nijenhuis bracket is a generalization of the Lie bracket on the space of multi-vectors. 

The terms in (\ref{eq:cfroytenberg}) that we have not calculated yet can be written, using (\ref{eq:rt2}), as
\begin{equation} \label{eq:rt3}
\left. \Big( ( {\cal{L}}_v a -{\cal{L}}_{u}b + \frac{1}{2} d (i_u b - i_v a) ) \Pi \Big) \right|^\mu = \Big( u^\nu (\partial_\nu b_\rho - \partial_\rho b_\nu) - v^\nu (\partial_\nu a_\rho - \partial_\rho a_\nu) + \frac{1}{2} \partial_\rho (ub - va) \Big) \Pi^{\rho \mu}.
\end{equation}
The Koszul bracket (\ref{eq:koszul}) can be further transformed in a following way
\begin{equation} \label{eq:rt4}
\left. [a,b]_\Pi \right|_\mu = \Pi^{\rho \nu} (b_\rho \partial_\nu a_\mu - a_\rho \partial_\nu b_\mu) + \partial_\nu \Pi^{\nu \rho} a_\rho b_\mu,
\end{equation}
while the remaining terms linear in $\Pi$ become
\begin{align} \notag \label{eq:rt5}
\left.( [-v, a\Pi]_L + [u, b \Pi]_L) \right|^\mu = &\  v^\nu( \partial_\nu a_\rho \Pi^{\mu \rho} + a_\rho \partial_\nu \Pi^{\mu \rho}) + a_\rho \Pi^{\rho \nu} \partial_\nu v^\mu \\
& - u^\nu (\partial_\nu b_\rho \Pi^{\mu \rho} + b_\rho \partial_\nu \Pi^{\mu \rho}) - b_\rho \Pi^{\rho \nu} \partial_\nu u^\mu.
\end{align}
Lastly, we write the expression for the Schouten-Nijenhuis bracket for bi-vectors
\begin{equation}
\left. [\Pi, \Pi]_S \right| ^{\mu \nu \rho} = \epsilon^{\mu \nu \rho}_{\alpha \beta \gamma} \Pi^{\sigma \alpha} \partial_\sigma \Pi^{\beta \gamma},
\end{equation}
where 
\begin{equation}
\epsilon^{\mu \nu \rho}_{\alpha \beta \gamma} = 
\begin{vmatrix}
\delta^\mu_\alpha & \delta^\nu_\beta & \delta^\rho_\gamma \\ 
\delta^\nu_\alpha & \delta^\rho_\beta & \delta^\mu_\gamma \\
\delta^\rho_\alpha & \delta^\mu_\beta & \delta^\nu_\gamma
\end{vmatrix}\ \ .
\end{equation}
Thus, we get
\begin{equation}
\left. ([\Pi, \Pi]_S(a,b,.)) \right|^\mu = 2 R^{\mu \nu \rho} a_\nu b_\rho,
\end{equation}
where $R^{\mu \nu \rho}$ is the flux defined in (\ref{eq:QRdef}).

Combining the previously obtained terms, we obtain the expression for the generalized current corresponding to the Roytenberg bracket twisted by the non-commutativity parameter as a bi-vector
\begin{equation} \label{eq:roytroyt}
[(u,a),(v,b)]_R = (w,c),
\end{equation}
where  $w^\mu$ and $c_\mu$ are equal to $\bar{w}^\mu$ and $\bar{c}_\mu$, defined in (\ref{eq:defw}) and (\ref{eq:defc}),  respectively, provided that  $\Pi^{\mu \nu} = \kappa \theta^{\mu \nu}$.

Integrating the previous equation over $\sigma$ and $\bar{\sigma}$, we see that charges satisfy
\begin{equation} \label{eq:qrbr}
\{ Q_{R(u,a)}, Q_{R(v,b)} \} = - Q_{R[(u,a),(v,b)]_R}.
\end{equation}
The bases in which these generalized currents have been defined are mutually T-dual (\ref{eq:can}). This means that the generalized currents also transform into each other 
\begin{equation}
J_{C (u,a)} \leftrightarrow J_{R (v,b)} \, ,
\end{equation}
provided that we swap also coefficients $u^\mu \leftrightarrow b_\mu$, $a_\mu \leftrightarrow v^\mu$. We say that two types of brackets, one obtained by twisting the Courant bracket by a 2-form $2 B_{\mu \nu}$, another obtained by twisting the Courant bracket by a bi-vector $\Pi^{\mu \nu}$, are mutually T-dual, as long as the aforementioned 2-form $B_{\mu \nu}$ is T-dual to the bi-vector $\Pi^{\mu \nu}$. 

In \cite{Tdualiso} it has been proposed that T-duality can be understood as the isomorphism $\varphi$ between two Courant algebroids \cite{courant1, gualtieri}. The relations connecting coefficients of two sets of generalized currents (\ref{eq:coeffc}) can in fact be rewritten as
\begin{equation} \label{eq:varphidef}
\varphi (u,a) = (u-\kappa \theta a, 2 B u + (G_E^{-1} G) a) \, ,
\end{equation}
which can be interpreted as the isomorphism $\varphi (u,a) = (v,b)$ between two Courant algebroids with the trivial bundles over a point and with the twisted Courant and Roytenberg brackets as brackets that act on the Cartesian product of sections of these bundles, as well as the natural inner product $\langle.,.\rangle$ between generalized vectors, given by
\begin{equation} \label{eq:isosp}
\langle(u,a),(v,b)\rangle = \frac{1}{2}( ub + va )\, .
\end{equation}
In order for $\varphi$ to be the isomorphism that corresponds to T-duality, it has to satisfy the following conditions:
\begin{equation} \label{eq:isoTdual}
\langle\varphi(u,a), \varphi(v,b)\rangle = \langle(u,a),(v,b)\rangle \, , \ \ [\varphi(u,a), \varphi(v,b)]_C = \varphi \Big( [ (u,a), (v,b) ]_R \Big) \, .
\end{equation}
To prove that the first condition is satisfied, using (\ref{eq:varphidef}), we obtain
\begin{align} 
\langle\varphi(u,a), \varphi(v,b)\rangle &= \langle(u-\kappa \theta a, 2 B u + (G_E^{-1} G) a), (v-\kappa \theta b, 2 B v + (G_E^{-1} G) b)\rangle  \\ \notag
& = \frac{1}{2} \Big(2 B_{\mu \nu} u^\mu v^\nu +2 \kappa (B\theta)_\mu^{\ \nu} v^\mu a_\nu + (G_E^{-1} G)^\mu_{\ \nu} b_\mu u^\nu - \kappa (G_E^{-1} G)^\mu_{\ \nu} \theta^{\nu \rho} b_\mu a_\rho  \\ \notag
& + 2 B_{\mu \nu} v^\mu u^\nu + 2 \kappa (B\theta)_\mu^{\ \nu} u^\mu b_\nu + (G_E^{-1} G)^\mu_{\ \nu} a_\mu v^\nu - \kappa (G_E^{-1} G)^\mu_{\ \nu} \theta^{\nu \rho} a_\mu b_\rho  \Big) \\ \notag
& = \frac{1}{2}(u^\mu b_\nu + v^\mu a_\nu) \Big( (G_E^{-1} G)^\nu_{\ \mu} + 2\kappa (\theta B)_{\ \mu}^{\nu} \Big) \\ \notag  
&= \langle(u,a),(v,b)\rangle \, ,
\end{align}
where we have used the fact that $B_{\mu \nu}$ and $(G_E^{-1} G \theta)^{\mu \nu}$ are both antisymmetric, as well as
\begin{equation}
(G_E^{-1} G)^\mu_{\ \nu} + 2\kappa (\theta B)_{\ \nu}^{\mu} = \delta^\mu_{\ \nu} \, ,
\end{equation}
which is the identity easily obtained from (\ref{eq:geff}) and (\ref{eq:ncdef}). As for the second relation of (\ref{eq:isoTdual}), it can be shown by writing the relation (\ref{eq:qqc}) for $\varphi$-transformed coefficients
\begin{equation} \label{eq:Qfpom}
\{ Q_{C \varphi(u, a)}, { Q_{C  \varphi(v, b)} \}} =  - Q_{C[ \varphi(u,a), \varphi(v,b)]_C} \, .
\end{equation}
On the other hand, due to $Q_{C \varphi(u,a)} = Q_{R(u,a)}$, the terms on the right-hand sides of (\ref{eq:Qfpom}) and (\ref{eq:qrbr}) are equal. By equating them, one obtains 
\begin{equation}
Q_{R[ (u,a), (v,b)]_R} = Q_{C[ \varphi(u,a), \varphi(v,b)]_C} \, .
\end{equation}
Lastly, using (\ref{eq:varphidef}), we write the above relation in the form
\begin{equation}
Q_{C \varphi ([ (u,a), (v,b)]_R)} = Q_{C[ \varphi(u,a), \varphi(v,b)]_C} \, ,
\end{equation}
from which the second condition of (\ref{eq:isoTdual}) is easily read. 

Therefore, we have shown that the relations connecting two types of generalized currents (\ref{eq:varphidef}) define the isomorphism between two Courant algebroids, characterized by twisted Courant and Roytenberg bracket, that according to \cite{Tdualiso} is interpreted as T-duality. 

\section{Conclusion}
\setcounter{equation}{0}
In this paper, we used the T-dualization rules (\ref{eq:xydual}) for coordinates in the Lagrangian approach, and (\ref{eq:tdual}) for the canonical variables in the Hamiltonian approach. The relation for T-dual background fields (\ref{eq:gbcan}) stands in both approaches. These relations between the fields provide correct relations between the Courant and Roytenberg bracket.

The T-dualization rules we used, correspond exactly to Buscher's rules obtained in its original procedure \cite{buscher,buscher1,buscher2} when there is an Abelian group of isometries of coordinates $x^a$ along which one T-dualizes: $B(x^a)= B(x^a+b^a)$, $G(x^a)=G(x^a+b^a)$. In the Buscher procedure the symmetry is gauged and the new action is obtained. Integrating out the gauge fields from that action, one obtains the T-dual Lagrangian. From that, the T-dual transformation law between the T-dual coordinate $\sigma$-derivatives and the canonical momenta of the initial theory can be obtained $ \kappa y^\prime_\mu \cong \pi_\mu$. This is exactly the relation (\ref{eq:tdual}) in our paper. 

The most interesting case is when we try to perform the T-dualization along non-isometry directions $x^a$, such that background fields do depend on them. Then we should apply the generalized Buscher's procedure, developed in \cite{tdwcb,tddrugi}. In this case, the expression for the T-dual background fields (\ref{eq:gbcan}) remain the same but the argument of the T-dual background fields is not simply the T-dual variable $y_a$. It is the line integral $V$ that is a function of the world-sheet gauge fields $v_{+}^a$ and $v_{-}^a$, namely $V^a [v_{+},v_{-}]\equiv \int_{P}d\xi^\alpha
v^{a}_{\alpha} =\int_{P}(d\xi^{+} v^{a}_{+} +d\xi^{-} v^{a}_{-}) $. The expressions for gauge fields can be obtained by varying the Lagrangian with respect to gauge fields and the expression for the argument of background fields has the form $V^a = - \kappa \, \theta^{a b } y_b +
G^{-1 a b}_E \, {\tilde y}_b$, where ${\tilde y}_a$ is a double of T-dual variable $y_a$, which satisfy relations $\dot{\tilde y}_a = y^ \prime_a$ and ${\tilde
y}^\prime_a = \dot y_a$. Let us point out that in such case the T-dual theory becomes locally non-geometric because the argument of the background fields is the line integral.

For example, in case of the weakly curved background \cite{tdwcb} the initial theory is geometric and T-dual theory is non-geometric. In the initial theory the generalized current algebra gives rise to the twisted Courant bracket. However, in the T-dual theory, the presence of double variable ${\tilde y}_a$,  makes the calculation of T-dual current algebra much more complicated. It is hard to believe that such a bracket or its corresponding self T-duality version will be equivalent to the Roytenberg one. Therefore, in case of non-geometric theories, one might expect some new form of brackets.

Next, we introduced the T-duality in the same phase space, that we call self T-duality. It interchanges the momenta and coordinate $\sigma$-derivatives, as well as the background fields with the T-dual ones. The Hamiltonian was expressed in terms of currents $j_{\pm \mu}$ and metric tensor $G_{\mu \nu}$, as well as in terms of its T-dual currents $l_\pm^\mu$ and T-dual metric tensor ${^\star G}^{\mu \nu} = (G_E^{-1})^{\mu \nu}$. We considered two types of generalized currents, $J_{C(u,a)}$ and $J_{R(v,b)}$, that generalize currents $j_{\pm \mu}$ and $l^\mu_\pm$ respectively. The suitable basis for the current $J_{C(u,a)}$ consists of coordinate $\sigma$-derivatives $x^{\prime \mu}$ and the auxiliary currents $i_\mu = \pi_\mu + 2 \kappa B_{\mu \nu} x^{\prime \nu}$, and for the current $J_{R(v,b)}$, it consists of momenta $\pi_\mu$ and auxiliary currents $k^\mu = \kappa x^{\prime \mu} + \kappa \theta^{\mu \nu} \pi_\nu$. These bases transform into each other under the self T-duality (\ref{eq:can}).

In this paper, we obtained two types of brackets, extracted from the generalized current Poisson bracket algebra. We have shown that one of them is equal to the twisted Courant bracket, while the other equals the Roytenberg bracket. The former can be obtained by twisting the Courant bracket by a 2-form, in our paper $2B_{\mu \nu}$, resulting in the appearance of $H-$flux in generators algebra. The latter bracket can be obtained by twisting the Courant bracket by a bi-vector $\Pi^{\mu \nu}$, 
resulting in the appearance of $Q-$ and $R-$fluxes, but not $H-$flux, in generators algebra. Since bases in which generalized currents are defined are mutually T-dual, we conclude that the brackets are mutually T-dual, when the bi-vector $\Pi^{\mu \nu}$  equals to the non-commutativity parameter $\kappa \theta^{\mu \nu}$.

We find these results important in itself. Both the Courant and the Roytenberg bracket are well understood mathematical structures. Relation between them and T-duality has a potential to help understand the T-duality better. Moreover, by analyzing characteristics of these brackets we can examine how certain aspects of the mutually T-dual theories relate to each other. 

Suppose we turn off all the fluxes. That is equivalent to setting $B_{\mu \nu} = 0$ and $\Pi^{\mu \nu} = 0$, which reduce the auxiliary currents to canonical momentum and coordinate $\sigma$ derivative: $i_\mu \to \pi_\mu$ and $k^\mu \to \kappa x^{\prime \mu}$. The generalized currents now reduce to $J_{C(u,a)} = u^\mu \pi_\mu + a_\mu \kappa x^{\prime \mu}$ and $J_{R(v,b)} = v^\mu \pi_\mu + \kappa  b_\mu x^{\prime \mu}$. It is easy to verify that these currents remain invariant under exchange of momenta and winding numbers, provided that we also change the coefficients in the particular way $J_{C(u,a)} \leftrightarrow \tilde{u}_\mu \kappa x^{\prime \mu} + \tilde{a}^\mu \pi_\mu = J_{C(\tilde{a},  \tilde{u})}$. Therefore, we conclude that these currents are T-dual to themselves. They give rise to the Courant bracket, the untwisted one, which does not contain any fluxes.

It is interesting that both charges $Q_{C(u,a)}$ and $Q_{R(v,b)}$ can be expressed as the self T-dual symmetry generators in the form
\begin{equation} \label{eq:GGG}
{\cal G}=\int d\sigma\Big[
\xi^\mu\pi_\mu+\widetilde\Lambda_\mu \kappa x^{\prime\mu}
\Big].
\end{equation}
It is easy to show that if we define the new gauge parameter $\Lambda_\mu = \widetilde\Lambda_\mu + 2 B_{\mu \nu} \xi^\nu$, the generators (\ref{eq:GGG}) are charges $Q_{C(\xi, \Lambda)}$; if we define $\tilde \xi^\mu = \xi^\mu + \kappa \theta^{\mu \nu} \widetilde\Lambda_\nu$ the generators are charges $Q_{R(\tilde{\xi}, \widetilde\Lambda)}$. Momenta $\pi_\mu$ are generators of general coordinate transformations and $x^{\prime \mu}$ generators of local gauge transformations $\delta_{\widetilde \Lambda} B_{\mu \nu} = \partial_\mu \widetilde\Lambda_\nu - \partial_\nu \widetilde\Lambda_\mu$, while $\xi^\mu$ and $\widetilde \Lambda_\mu$ are their corresponding parameters. These generators were studied in \cite{dualsim}, where it was shown that general coordinate transformations are T-dual to gauge transformations.

\end{document}